\documentclass[aps,prd,showpacs,preprintnumbers,twocolumn,nofootinbib]{revtex4}
\usepackage[dvips]{graphicx}
\usepackage{bm,latexsym,amsmath,amssymb,amsfonts}
\newcommand*{\cP}{{\cal P}}
\begin{document}

\title{
The spectrum of gravitational waves in Randall-Sundrum braneworld cosmology
}

\author{Tsutomu~Kobayashi}
\email{tsutomu@tap.scphys.kyoto-u.ac.jp}
\author{Takahiro~Tanaka}
\email{tama@scphys.kyoto-u.ac.jp}

\affiliation{
Department of Physics, Kyoto University, Kyoto 606-8502, Japan 
}

\begin{abstract}
We study the generation and evolution of gravitational waves
(tensor perturbations) in the context of Randall-Sundrum
braneworld cosmology.
We assume that the initial and final stages of
the background cosmological model are given by
de Sitter and Minkowski phases, respectively,
and they are connected smoothly by a radiation-dominated phase.
This setup allows us to discuss
the quantum-mechanical generation of the perturbations
and to see the final amplitude of the well-defined zero mode.
Using the Wronskian formulation,
we numerically compute the power spectrum of
gravitational waves,
and find that the effect of
initial vacuum fluctuations in the Kaluza-Klein modes is subdominant,
contributing not more than 10\% of the total power spectrum.
Thus it is confirmed that
the damping due to the Kaluza-Klein mode generation
and the enhancement due to the modification of the background
Friedmann equation are the two dominant effects,
but they cancel each other,
leading to the same spectral tilt as the standard four-dimensional result.
Kaluza-Klein gravitons that escape from the brane
contribute to the energy density of the dark radiation at late times.
We show that a tiny amount of the dark radiation is generated
due to this process.
\end{abstract}


\preprint{KUNS-1995}

\maketitle

\section{Introduction}

Cosmological inflation predicts
the gravitational wave background arising
due to quantum fluctuations in the graviton field.
Gravitational wave fluctuations are
stretched beyond the horizon radius by rapid expansion during inflation,
and at a later stage they come back inside the horizon
possibly with rich information on
the early universe and hence on high energy physics.
Though yet undetected, gravitational waves will
provide us with a powerful tool to probe fundamental physics
in near future~\cite{Maggiore:1999vm}.

Motivated by string theory,
recently braneworld scenarios have attached much attention,
in which our four-dimensional universe is realized
as a brane embedded
in a higher dimensional bulk spacetime~\cite{Maartens:2003tw}.
Among them, the Randall-Sundrum model~\cite{Randall:1999ee, Randall:1999vf}
is of particular interest
because it includes nontrivial gravitational dynamics
despite rather a simple construction.
In the Randall-Sundrum type II model~\cite{Randall:1999vf}
with a single brane embedded in an anti-de Sitter (AdS) bulk,
although the fifth dimension extends infinitely,
the warped structure of the bulk geometry
results in the recovery of four-dimensional general relativity
on the brane at scales larger than the bulk curvature scale $\ell$
or at low energies~\cite{Garriga:1999yh, Shiromizu:1999wj}.
In order to reveal five-dimensional effects
particular to the braneworld scenario,
we have to focus on the scales smaller than $\ell$,
and for this purpose cosmological perturbations from
inflation~\cite{CP, Langlois:2000ns, Gorbunov:2001ge, Kobayashi:2003cn, Hiramatsu:2003iz, Hiramatsu:2004aa, Ichiki:2003hf, Ichiki:2004sx, Easther:2003re, Battye:2003ks, Battye:2004qw, Cartier:2005br, Tanaka:2004ig, Kobayashi:2004wy, Kobayashi:2005jx}
will be quite useful
for the reason mentioned above.

Gravitational waves from inflation on the brane
were first studied by Langlois \textit{et al.}~\cite{Langlois:2000ns},
under an assumption that inflation is exactly described by de Sitter spacetime.
In this special case, the perturbation equation is separable
and analytically solvable.
A toy model called
the ``junction model''~\cite{Gorbunov:2001ge, Kobayashi:2003cn}
is an extended version of
the pure de Sitter braneworld,
which allows a sudden change of the Hubble parameter $H$
by joining two maximally symmetric (i.e., de Sitter or Minkowski) branes
at some time.
Later, the junction model is extended to a
more general inflation model with a smooth expansion rate~\cite{Kobayashi:2005jx}.
To make the cosmological model more realistic,
one should take into account the radiation-dominated phase
that follows after inflation,
and,
at least in the low energy regime ($\ell H \ll 1$),
corrections to the evolution of gravitational waves
are shown to be small~\cite{Tanaka:2004ig, Kobayashi:2004wy}
(see also Refs.~\cite{Easther:2003re, Battye:2003ks, Battye:2004qw, Cartier:2005br}).
In a much more general and interesting case, i.e.,
in the high energy ($\ell H\gg 1$) radiation-dominated phase,
the perturbation equation no longer has a separable form
and hence one cannot even define
a ``zero mode'' and ``Kaluza-Klein modes'' without ambiguity.
To understand the evolution of gravitational waves
in that regime, numerical studies have been done
by Hiramatsu \textit{et al.}~\cite{Hiramatsu:2003iz, Hiramatsu:2004aa}
and by Ichiki and Nakamura~\cite{Ichiki:2003hf, Ichiki:2004sx}.
Their results give us a lot of implications, for example,
on the damping nature of the gravitational wave amplitude
due to the Kaluza-Klein mode generation,
but the initial condition they adopt is naive,
neglecting initial quantum fluctuations in the Kaluza-Klein modes.
Hence, its validity is open to question.

The goal of the present paper is to clarify
the late time power spectrum of gravitational waves
in the Randall-Sundrum brane cosmology,
evolving through the radiation-dominated stage
after their generation during inflation.
We closely follow the same line in our previous work~\cite{Kobayashi:2005jx},
in which, using the Wronskian formulation, we have formulated
a numerical scheme for the braneworld cosmological perturbations.
The initial condition in our analysis
is imposed quantum-mechanically,
and therefore we will be able to obtain
a true picture of the generation and evolution of
gravitational perturbations in the braneworld.

This paper is organized as follows.
In the next section, we start with giving
the background cosmological model
and summarize basic known results
concerning the gravitational wave
mode functions in the de Sitter and Minkowski braneworlds.
In Section~III, we describe the Wronskian formulation to
obtain the power spectrum of gravitational waves,
and then we show our numerical results in Sec.~IV.
In Section~V we discuss an amount of the dark radiation
generated due to excitation of Kaluza-Klein modes.
Finally we conclude in Sec.~VI.

\section{Preliminaries}

\subsection{The background model}

Now we describe a model for the background.
We shall work in the cosmological setting
of the Randall-Sundrum braneworld,
and so the bulk is given by a five-dimensional AdS spacetime.
The AdS metric in the Poincor\'{e} coordinates is
\begin{eqnarray}
ds^2=\frac{\ell^2}{z^2}(-dt^2+\delta_{ij}dx^idx^j+dz^2),
\label{AdS_Poincare}
\end{eqnarray}
where $\ell$ is the bulk curvature scale and constrained by
table-top experiments as $\ell \lesssim 0.1$ mm~\cite{Long:2002wn}.
A cosmological brane moves in this static bulk, the trajectory of which
is given by $z=z(t)$.
The scale factor of the universe
is related to the position of the brane as $a(t)=\ell/z(t)$.

\begin{figure}[t]
  \begin{center}
    \includegraphics[keepaspectratio=true,height=100mm]{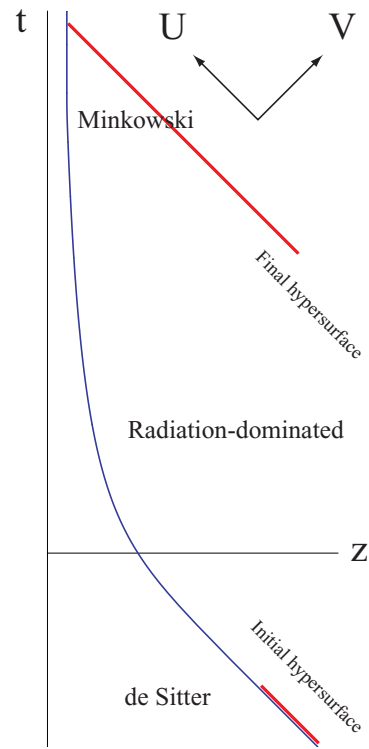}
  \end{center}
  \caption{Brane trajectory in static coordinates.}%
  \label{fig:trajectory.eps}
\end{figure}

We consider the following cosmological model on the brane.
The initial stage of the model is given by de Sitter inflation
with a constant Hubble parameter $H=H_i$,
which is smoothly connected to the radiation-dominated phase.
(In order to join the two phases smoothly,
the brane is not exactly de Sitter at the very last stage of inflation.)
In the radiation stage the scale factor evolves subject to
the modified Friedmann equation~\cite{FRW_brane}
\begin{eqnarray}
H^2=\frac{\rho_r}{3M_{{\rm Pl}}^2}\left(1+\frac{\rho_r}{2\sigma}\right),
\end{eqnarray}
where $\rho_r$ is the radiation energy density and $\sigma=6M_{{\rm Pl}}^2/\ell^2$
is the tension of the brane.
Since the conventional conservation law holds on the brane,
we have $\rho_r\propto a^{-4}$.
Thus, in terms of the proper time $\tau$ on the brane we obtain
\begin{eqnarray}
a(\tau)=a(\tau_1)\left[
\left(\frac{\tau}{\tau_1}\right)^2+2c\left(\frac{\tau}{\tau_1}\right)-2c
\right]^{1/4},
\end{eqnarray}
where $\tau_1$ is a fiducial time,
$c:=\sqrt{1+[\rho_r(\tau_1)/2\sigma]}-1$,
and $\rho_r(\tau_1)/\sigma=\ell^2/8\tau_1^2$.
After a period of time the energy scale of the universe becomes
sufficiently low, and the radiation-dominated phase is then smoothly connected
to the Minkowski phase.
This artificial connection will not cause any unexpected problems
on our final result (i.e., power spectra of gravitational waves)
because at the end of the radiation stage
the brane universe is already in the low energy regime.
Just for simplicity we assume
that the Minkowski brane is located at $z=\ell$.
Namely, the scale factor is normalized so that
$a=a_0=1$ when the universe ceases expanding.
The motion of the brane is shown in Fig.~\ref{fig:trajectory.eps}.

\subsection{Minkowski and de Sitter braneworlds}

Let us consider tensor perturbations in AdS spacetime
bounded by a brane.
We can decompose the graviton field into a zero mode and
Kaluza-Klein (KK) modes without ambiguity
when the brane is maximally symmetric.
This is the reason why the initial and final stages of the background model 
are given by the de Sitter and Minkowski phases, respectively.

We write the perturbed metric as
\begin{eqnarray}
ds^2=\frac{\ell^2}{z^2}\left[-dt^2+(\delta_{ij}+h_{ij})dx^idx^j+dz^2\right],
\label{perturbed_Poincare}
\end{eqnarray}
where $h_{ij}$ is the transverse-traceless metric perturbation.
As usual we
decompose it into the spatial Fourier modes as
\begin{eqnarray}
h_{ij}=\frac{\sqrt{2}}{(2\pi M_5)^{3/2}}\int d^3k~\phi_{{\bf k}}e^{i{\bf k}\cdot{\bf x}}e_{ij}.
\end{eqnarray}
Here $M_5$ is the fundamental mass scale which is related to
the four-dimensional Planck mass $M_{{\rm Pl}}$ by
$\ell(M_5)^3=M_{{\rm Pl}}^2$.
The prefactor $\sqrt{2}/(M_5)^{3/2}$ is chosen so that
the kinetic term for $\phi_{{\bf k}}$ in the effective action is
canonically normalized.
From now on we suppress the subscript ${\bf k}$.

For the analysis of perturbations from the Minkowski brane,
the above Poincar\'{e} coordinate system will be best suited,
and the perturbation equation is
\begin{eqnarray}
\left(\frac{\partial^2}{\partial t^2}+k^2-\frac{\partial^2}{\partial z^2}
+\frac{3}{z}\frac{\partial}{\partial z}\right) \phi=0,
\label{KG_Poincare}
\end{eqnarray}
subject to the boundary condition
\begin{eqnarray}
\partial_z\phi\big|_{z=\ell}=0.
\end{eqnarray}
Now it is easy to find mode solutions of Eq.~(\ref{KG_Poincare}).
Going to quantum theory, the graviton field can be expanded
in terms of the zero mode and KK modes as
\begin{eqnarray}
\phi = \hat A_0 \varphi_0+\hat A_0^{\dagger} \varphi_0^*
+\int_0^{\infty} \!\!dm\left(\hat A_m \varphi_m+\hat A_m^{\dagger} \varphi_m^* \right),
\label{quantum_Min}
\end{eqnarray}
where $\hat A_n$ and $\hat A_n^{\dagger}$ ($n=0, m$) are
the annihilation and creation operators, respectively, of their corresponding modes.
The normalized zero mode function is given by
\begin{eqnarray}
\varphi_0(t)=\frac{1}{\sqrt{2k\ell}}e^{-ikt},
\end{eqnarray}
while the normalized KK mode function is
\begin{eqnarray}
\varphi_m(t,z)=\frac{1}{\sqrt{2\omega \ell^3}}e^{-i\omega t}u_m(z),
\end{eqnarray}
with
\begin{eqnarray}
u_m(z):=z^2\sqrt{\frac{m}{2}}
\frac{Y_1(m\ell)J_2(mz)-J_1(m\ell)Y_2(mz)}
{\sqrt{[Y_1(m\ell)]^2+[J_1(m\ell)]^2}},
\end{eqnarray}
and
\begin{eqnarray}
\omega=\sqrt{k^2+m^2}.
\end{eqnarray}
The normalization here is determined by the Wronskian conditions
\begin{eqnarray}
&&(\varphi_0 \cdot \varphi_0) = -(\varphi_0^*\cdot\varphi_0^*) = 1,
\label{Wronskian_condition_min}\\
&&(\varphi_{m} \cdot \varphi_{m'}) = - (\varphi_{m}^*\cdot\varphi_{m'}^*)
= \delta(m-m'),
\nonumber\\
&&(\varphi_0 \cdot \varphi_{m}) = (\varphi_{0}^*\cdot\varphi_{m}^*)
=0,
\nonumber\\
&&(\varphi_{n}\cdot\varphi_{n'}^*) = 0,
\qquad\mbox{for}\quad n,n'=0, m,
\nonumber
\end{eqnarray}
where the Wronskian is defined by~\cite{Gorbunov:2001ge}
\begin{eqnarray}
(X\cdot Y):=-2i\int^{\infty}_{\ell}dz\left(\frac{\ell}{z}\right)^3
\left(X\partial_tY^*-Y^*\partial_tX\right).
\end{eqnarray}

In the de Sitter braneworld we introduce
another set of coordinates $(\eta, \xi)$,
which is related to $(t, z)$ as
\begin{eqnarray}
t=\eta\cosh\xi+t_0,\quad z=-\eta\sinh\xi,
\end{eqnarray}
where $t_0$ is an arbitrary constant.
In $(\eta,\xi)$ frame
the de Sitter brane is located at a fixed coordinate position $\xi=\xi_b=$ constant,
and the Hubble parameter on the brane is given by $H_i=\ell^{-1}\sinh\xi_b$.
The perturbation equation
again has a separable form
\begin{eqnarray}
\left(\frac{\partial^2}{\partial\eta^2}-\frac{2}{\eta}\frac{\partial}{\partial\eta}+k^2
-\frac{\sinh^3\xi}{\eta^2}\frac{\partial}{\partial\xi}\frac{1}{\sinh^3\xi}
\frac{\partial}{\partial\xi}\right)\phi=0,
\label{KGeq_in_eta_xi}
\end{eqnarray}
subject to the boundary condition
\begin{eqnarray}
\partial_{\xi}\phi\big|_{\xi=\xi_b}=0.
\end{eqnarray}
Treating $\phi$ as an operator, the graviton field can be expanded as
\begin{eqnarray}
\phi=\hat a_0\phi_0+\hat a_0^{\dagger}\phi_0^*
+\int_{0}^{\infty}\!\!d\nu\left(
\hat a_{\nu}\phi_{\nu}+\hat a_{\nu}^{\dagger}\phi_{\nu}^*
\right),
\label{quantum_dS}
\end{eqnarray}
where $\hat a_n$ and $\hat a_n^{\dagger}$ ($n=0, \nu$) are
the annihilation and creation operators of each mode.
The explicit form of the normalized zero mode is
\begin{eqnarray}
\phi_0(\eta)=C(\ell H_i)\cdot
\frac{H_i}{\sqrt{2k\ell}}\left(
\eta-\frac{i}{k}\right) e^{-ik\eta},
\end{eqnarray}
with
\begin{eqnarray}
C(x):=\left[
\sqrt{1+x^2}+x^2\ln\left(\frac{x}{1+\sqrt{1+x^2}}\right)\right]^{-1/2},
\end{eqnarray}
and the KK mode functions are found in the form of
$\phi_{\nu}(\eta,\xi)=\psi_{\nu}(\eta)\chi_{\nu}(\xi)$,
where
\begin{eqnarray}
\psi_{\nu}(\eta) &=& \frac{\sqrt{\pi}}{2}\ell^{-3/2}
e^{-\pi\nu/2}(-\eta)^{3/2}H^{(1)}_{i\nu}(-k\eta),
\\
\chi_{\nu}(\xi) &=& C_1(\sinh\xi)^2
\left[
P^{-2}_{-1/2+i\nu}(\cosh\xi)\right.
\nonumber\\
&&\quad\left.-C_2
Q^{-2}_{-1/2+i\nu}(\cosh\xi) \right],
\end{eqnarray}
with
\begin{eqnarray}
C_1&=&\left[\left| \frac{\Gamma(i\nu)}{\Gamma(5/2+i\nu)}\right|^2\right.
\nonumber\\
&&\hspace{-3mm}
\left.+\left| \frac{\Gamma(-i\nu)}{\Gamma(5/2-i\nu)}
-\pi C_2\frac{\Gamma(i\nu-3/2)}{\Gamma(1+i\nu)}\right|^2\right]^{-1/2},
\\
C_2&=&
\frac{P^{-1}_{-1/2+i\nu}(\cosh\xi_b)}{Q^{-1}_{-1/2+i\nu}(\cosh\xi_b)}.
\end{eqnarray}
Note that the index $\nu(\geq 0)$ is related to the Kaluza-Klein mass as
\begin{eqnarray}
m^2=\left(\nu^2+\frac{9}{4}\right)H_i^2.
\end{eqnarray}
The normalization of the modes is determined by the Wronskian conditions
\begin{eqnarray}
&&(\phi_0 \cdot \phi_0) = -(\phi_0^*\cdot\phi_0^*) = 1,
\label{Wronskian_condition_dS}\\
&&(\phi_{\nu} \cdot \phi_{\nu'}) = - (\phi_{\nu}^*\cdot\phi_{\nu'}^*)
= \delta(\nu-\nu'),
\nonumber\\
&&(\phi_0 \cdot \phi_{\nu}) = (\phi_{0}^*\cdot\phi_{\nu}^*)
=0,
\nonumber\\
&&(\phi_{n}\cdot\phi_{n'}^*) = 0,
\qquad\mbox{for}\quad n,n'=0, \nu.
\nonumber
\end{eqnarray}
Here the Wronskian is written in $(\eta, \xi)$ frame as~\cite{Gorbunov:2001ge},
\begin{eqnarray}
(X\cdot Y):=-2i\int^{\infty}_{\xi_b}d\xi\frac{\ell^3}{\eta^2\sinh^3\xi}
\left(X\partial_{\eta}Y^*-Y^*\partial_{\eta}X\right).
\nonumber\\
\end{eqnarray}

\section{Wronskian formulation}

Due to the presence of an infinite tower of Kaluza-Klein modes,
cosmological perturbations in the braneworld have infinite degrees of freedom.
Instead of solving an initial value problem for such a system,
it would be better to use the Wronskian formulation
in order to take necessary degrees of freedom out of infinite information.
In the present case, we would like to know the final amplitude of the zero mode,
and therefore in fact what we need to do is solving
the (backward) evolution of a single degree
of freedom~\cite{Gorbunov:2001ge, Kobayashi:2003cn, Kobayashi:2005jx}.
Following the same line as our previous work~\cite{Kobayashi:2005jx},
we now explain how we compute
the amplitude of gravitational waves in the final Minkowski phase
using the Wronskian.

Using double null coordinates
\begin{eqnarray}
&u=t-z,\\
&v=t+z,
\end{eqnarray}
which will be convenient
for numerical calculations,
the metric~(\ref{AdS_Poincare})
can be rewritten in the form of
\begin{eqnarray}
ds^2=\frac{4\ell^2}{(v-u)^2}\left(
-dudv+\delta_{ij}dx^idx^j
\right).
\end{eqnarray}
The trajectory of the brane can be 
specified arbitrarily by 
\begin{eqnarray}
v=q(u).
\end{eqnarray}
By a further coordinate transformation
\begin{eqnarray}
&&U=u,\\
&&q(V)=v,
\end{eqnarray}
we obtain
\begin{eqnarray}
ds^2=\frac{4\ell^2}{[q(V)-U]^2}\left[
-q'(V)dUdV+\delta_{ij}dx^idx^j
\right],
\end{eqnarray}
where a prime denotes differentiation with respect to the argument. 
Now in the new coordinates the position of the brane is simply given by
\begin{eqnarray}
U=V.
\end{eqnarray}
We will use this coordinate system for actual numerical calculations.

The induced metric on the brane is
\begin{eqnarray}
ds_b^2=\frac{4\ell^2}{[q(V)-V]^2}\left[
-q'(V)dV^2+\delta_{ij}dx^idx^j
\right],
\end{eqnarray}
from which we can read off the scale factor $a$ and
the proper time $\tau$, respectively, as
\begin{eqnarray}
&&a = \frac{2\ell}{q(V)-V},\label{a-q}\\
&&d\tau  = a\sqrt{q'(V)} ~dV,\label{eta-V}
\end{eqnarray}
and hence the Hubble parameter on the brane
is written as
\begin{eqnarray}
\ell H = \frac{1}{2\sqrt{q'(V)}}[1-q'(V)],
\end{eqnarray}
or equivalently
\begin{eqnarray}
q'(V)=\left(\sqrt{1+\ell^2H^2}-\ell H\right)^2.
\label{q-H}
\end{eqnarray}
Given the Hubble parameter as a function of $\tau$,
one can integrate Eqs.~(\ref{eta-V}) and~(\ref{q-H})
with the aid of $da/d\tau=aH$
to obtain $q$ as a function of $V$.
If the Hubble parameter on the brane
is constant in time, we have $q'=$ constant.
Especially, $q'=1$ in the Minkowski phase.

The Klein-Gordon-type equation for a gravitational wave perturbation $\phi$
in the $(U, V)$ coordinates
reduces to
\begin{eqnarray}
\!
\left[4\partial_U\partial_V
+\frac{6}{q(V)-U}(\partial_V-q'(V)\,\partial_U)
+q'(V)\,k^2\right]\phi=0,\!\!\!
\cr
\label{KG_in_UV}
\end{eqnarray}
supplemented by the boundary condition
\begin{eqnarray}
[\partial_U-\partial_V]\phi\bigr\vert_{U=V}=0.
\label{bc}
\end{eqnarray}
The expression for the Wronskian evaluated on a constant $V$
hypersurface is given by 
\begin{eqnarray}
(X\cdot Y)={2i}\!\int^{V}_{-\infty}\!\!\!
 dU\! \left[\frac{2\ell}{q(V)-U}\right]^{3}\!
(X\partial_UY^*\!-Y^*\partial_U X),\cr
\end{eqnarray}
which is independent of the choice of the hypersurface.

As explained in Section~II, our cosmological model is composed of
the de Sitter inflationary phase followed by
the radiation-dominated epoch, which is connected smoothly to
the final Minkowski phase.
In the initial de Sitter phase, the graviton field can be expanded
as Eq.~(\ref{quantum_dS}),
while in the final Minkowski phase it can be expanded
as Eq.~(\ref{quantum_Min}).
We assume that initially the gravitons are in the de Sitter invariant
vacuum state annihilated by $\hat a_0$ and $\hat a_{\nu}$,
\begin{eqnarray}
\hat a_0|0\rangle=\hat a_{\nu}|0\rangle=0.
\end{eqnarray}
The expectation value of the squared amplitude of
the zero mode in the final stage is
\begin{eqnarray*}
\langle 0|
\left(\varphi_0 \hat A_0+\varphi^*_0\hat A_0^{\dagger}\right)^2
|0\rangle
&=&|\varphi_0|^2\langle 0|\left(1+2\hat A_0^{\dagger}\hat A_0\right)|0\rangle
\\
\qquad&&+~\mbox{oscillating part}
\\
&\simeq&\frac{1}{k\ell}N_f,
\end{eqnarray*}
where $N_f:=\langle 0|\hat A_0^{\dagger}\hat A_0|0\rangle$
is the number of created zero mode gravitons.
Here we used the commutation relation
$\left[\hat A_0, \hat A_0^{\dagger}\right]=1$
and assumed that $N_f\gg 1$.
The final power spectrum is then given by
\begin{eqnarray}
\cP(k)&:=&\frac{4\pi k^3}{(2\pi)^3}\frac{2}{(M_5)^3}\cdot
\frac{1}{k\ell}N_f
\nonumber\\
&=&\frac{k^2}{\pi^2M_{{\rm Pl}}^2}N_f.
\label{def_ps}
\end{eqnarray}
The operator $\hat A_0$ can be projected out
by making use of the Wronskian relations.
Noting that the Wronskian is constant in time,
we have
\begin{eqnarray}
\hat A_0&=&(\phi\cdot \varphi_0)_f=(\phi\cdot\Phi)
\nonumber\\
&=&(\phi_0\cdot\Phi)_i\hat a_0
+\int d\nu(\phi_{\nu}\cdot\Phi)_i\hat a_{\nu}+~\mbox{h.c.},
\end{eqnarray}
where $\Phi$ is a solution of the Klein-Gordon equation~(\ref{KG_in_UV})
whose final configuration is the zero mode function $\varphi_0$
in the Minkowski phase,
and
subscript $f$ and $i$ denote the quantities
evaluated on the final and initial hypersurfaces, respectively.
Thus we clearly see that final zero mode gravitons
are created from the vacuum fluctuations
both in the initial zero mode and in the KK modes:
\begin{eqnarray}
N_f=|(\phi_0^*\cdot \Phi)_i|^2+\int d\nu|(\phi_{\nu}^*\cdot \Phi)_i|^2.
\end{eqnarray}
Correspondingly, the power spectrum~(\ref{def_ps})
can be written as a sum of the two contributions:
\begin{eqnarray}
\cP=\cP_0+\cP_{{\rm KK}},
\end{eqnarray}
where
\begin{eqnarray}
\cP_0&:=&\frac{k^2}{\pi^2M_{{\rm Pl}}^2}|(\phi_0^*\cdot \Phi)_i|^2,
\\
\cP_{{\rm KK}}&:=&\frac{k^2}{\pi^2M_{{\rm Pl}}^2}\int d\nu|(\phi_{\nu}^*\cdot \Phi)_i|^2.
\end{eqnarray}

\section{Spectrum of gravitational waves}

De Sitter inflation on the brane predicts
the flat primordial spectrum~\cite{Langlois:2000ns}
\begin{eqnarray}
\delta_T^2:=\frac{2C^2(\ell H_i)}{M_{{\rm Pl}}^2}\left(\frac{H_i}{2\pi}\right)^2.
\label{primordial_specturm}
\end{eqnarray}
During inflation the gravitational wave perturbations are
stretched to super-horizon scales,
and then they stays constant until horizon reentry,
with their amplitude given by Eq.~(\ref{primordial_specturm}).
($\phi =$ constant is a growing solution of Eq.~(\ref{KG_in_UV})
in the limit $k^2\to 0$.)
The primordial spectrum of gravitational waves from non-de Sitter inflation
was studied extensively in~\cite{Kobayashi:2005jx}, and so in this paper
we concentrate on the simple case where inflation is given by the exact de
Sitter model.

For long wavelength modes with $k\ll k_*$, where
\begin{eqnarray}
k_*:=a_*H_*=a_*/\ell
\end{eqnarray}
labels the mode that reenters the horizon when $\ell H=1$,
the amplitude will decay as $h_{\mu\nu}\propto a^{-1}$
after horizon reentry, because gravity on the brane
is basically described by four-dimensional general relativity
in the low energy regime ($\ell H\ll 1$).
In fact, it is explicitly shown that
leading order corrections to the cosmological evolution
of gravitational waves are suppressed by $\ell^2$ and $\ell^2\ln \ell$
at low energies~\cite{Tanaka:2004ig, Kobayashi:2004wy}.
In particular, modes with $k<k_0$, where
\begin{eqnarray}
k_0:=a_0H_0=H_0,\label{k0a0H0}
\end{eqnarray}
and $H_0$ is the Hubble parameter evaluated at
the end of the radiation-dominated phase
(i.e., just before the Minkowski phase)\footnote{To be
more precise, the scale factor in Eq.~(\ref{k0a0H0})
should be replaced by a bit smaller one than $a_0=1$,
because $a_0$ is defined as the scale factor
in the Minkowski phase (where the Hubble parameter vanishes).
However, as the smooth period connecting the radiation and Minkowski phases
is taken to be very short, this point is not important.},
reenter the ``horizon'' in the final Minkowski phase,
and then they begin to oscillate (but their amplitude will not decay).
As a result, the mean-square vacuum fluctuations of such modes become
half of the initial value, leading to
\begin{eqnarray}
\cP = \frac{\delta_T^2}{2},\quad \mbox{for}~ k<k_0.
\end{eqnarray}
In Ref.~\cite{Gorbunov:2001ge}, the same spectrum was obtained
for scales larger than the AdS and horizon scales
by using the ``junction model'', in which an instantaneous transition
from a de Sitter to a Minkowski brane was assumed.
For the reason mentioned above, modes with $k_0< k\ll k_*$
have the standard spectrum,
\begin{eqnarray}
\cP = \frac{\delta_T^2}{2}\left(\frac{k}{k_0}\right)^{-2}.
\label{sp_standard}
\end{eqnarray}

Something nontrivial may happen to
gravitational waves with $k\gtrsim k_*$.
If the effects of mode mixing are neglected,
only the modification of the background expansion rate
alters the spectrum for these short wavelength modes to
\begin{eqnarray}
\tilde\cP \simeq \frac{\delta_T^2}{2}\left(\frac{k}{k_*}\right)^{-2/3}
\left(\frac{k_*}{k_0}\right)^{-2}.
\label{spectrum_incorrect}
\end{eqnarray}
However, this evaluation will not be correct
because mode mixing is expected to be efficient at high energies.

Our procedure to obtain a correct power spectrum is as follows.
We solve the perturbation equation~(\ref{KG_in_UV})
with its boundary condition set to be $\Phi(U, V_f)=\varphi_0$
on the final hypersurface.
The numerical backward evolution scheme
we use here is the same as that used in our previous work~\cite{Kobayashi:2005jx},
and the detailed description of the scheme is found there.
After obtaining the configuration on the initial hypersurface,
we evaluate the Wronskian to get the power spectrum.

We performed numerical calculations
for three different values of the inflationary Hubble parameter,
$\ell H_i=10$, $42$, and $100$.
The radiation-dominated phase is terminated
when the Hubble parameter decreases down to
$H\simeq 0.03/\ell=:H_0$, and then it is connected smoothly
to the Minkowski phase,
so that we can see the amplitude of the well-defined zero mode.
The numbers of grids are 90000 in the $U$-direction
and 12000 in the $V$-direction,
and the grid separation is chosen to be $\sim 0.005\times \ell$.
Integration over the KK index $\nu$ is performed up to $\nu=25$
with an equal grid spacing of $0.05$.

\begin{figure}[t]
  \begin{center}
    \includegraphics[keepaspectratio=true,height=150mm]{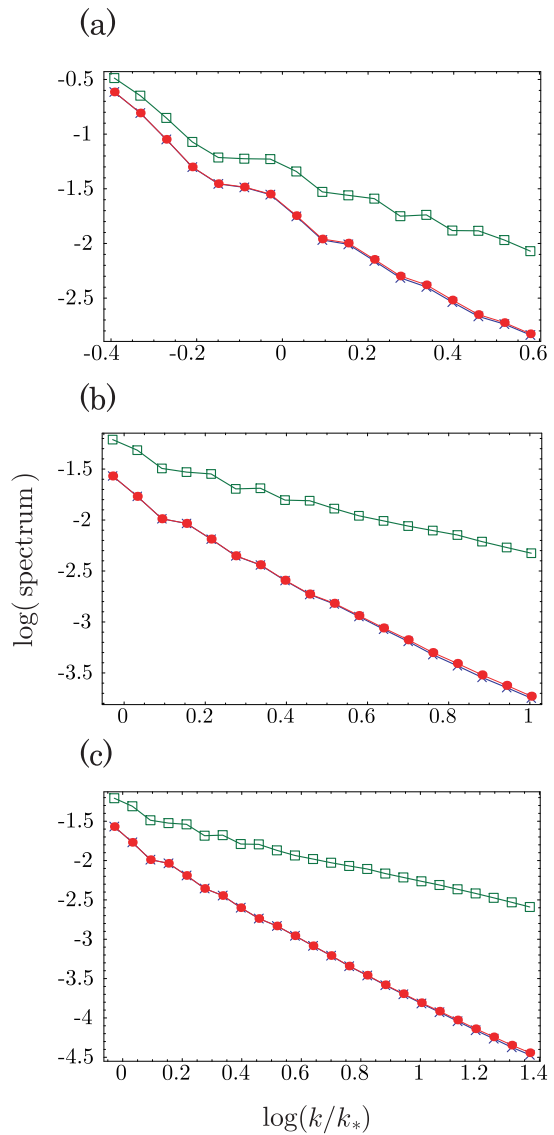}
  \end{center}
  \caption{Power spectra of gravitational waves from inflation
  with $\ell H_i=10$ (a), $42$ (b), and $100$ (c), normalized by $\delta_T^2/2$.
  The total power spectrum is shown by red circles, while blue crosses
  represent the contribution only from the initial zero mode.
  Green squires indicate results from
  basically four-dimensional calculations, only including
  the effect of the modification of the background expansion rate.
  }%
  \label{fig:spectrum.eps}
\end{figure}

\begin{figure}[t]
  \begin{center}
    \includegraphics[keepaspectratio=true,height=50mm]{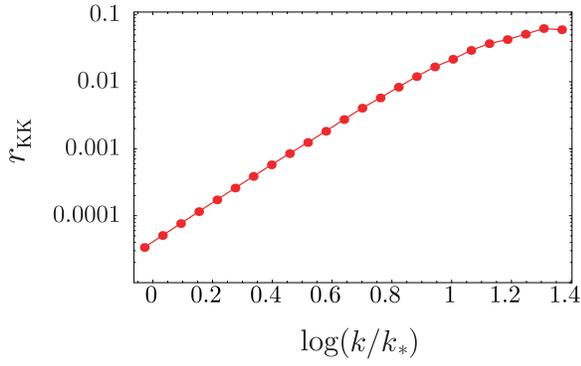}
  \end{center}
  \caption{Contribution of initial Kaluza-Klein fluctuations, $r_{{\rm KK}}$,
  for the model with $\ell H_i=100$.}
  \label{fig:kk.eps}
\end{figure}

The power spectra of gravitational waves are shown
in Fig.~\ref{fig:spectrum.eps}.
Assuming that the power spectrum is of the form
\begin{eqnarray}
\cP = A\left(\frac{k}{k_0}\right)^n,
\label{psp1}
\end{eqnarray}
we find that irrespective of the inflationary energy scale,
the parameters are approximately given by
\begin{eqnarray}
A&\simeq&\frac{\delta_T^2}{2},
\\
n&\simeq&-2.
\label{psp3}
\end{eqnarray}
Namely, we have the same spectrum as
the standard one [Eq.~(\ref{sp_standard})]
even for short wavelength modes with $k\gtrsim k_*$.
As is shown in Fig.~\ref{fig:kk.eps},
the contribution of the vacuum fluctuations
in the initial KK modes to the final spectrum,
\begin{eqnarray}
r_{{\rm KK}}(k):=\frac{\cP_{{\rm KK}}}{\cP_0+\cP_{{\rm KK}}},
\end{eqnarray}
never exceeds 10\% so far as the present calculations are concerned,
and hence
it gives a subdominant effect.
On the other hand, the excitation of KK modes
suppresses the amplitude of the gravitational waves
relative to Eq.~(\ref{spectrum_incorrect}),
and our result implies that the effect of
the modification of the background Friedmann equation
compensate this suppression,
leading to approximately the same spectral tilt as that in conventional
four-dimensional cosmology.
This is consistent with the numerical study by
Hiramatsu \textit{et al}.~\cite{Hiramatsu:2004aa}, in which
they assume the initial configuration in the bulk to be
a de Sitter zero mode and obtain $\cP_0 \propto k^{-2}$.

\begin{figure}[t]
  \begin{center}
    \includegraphics[keepaspectratio=true,height=50mm]{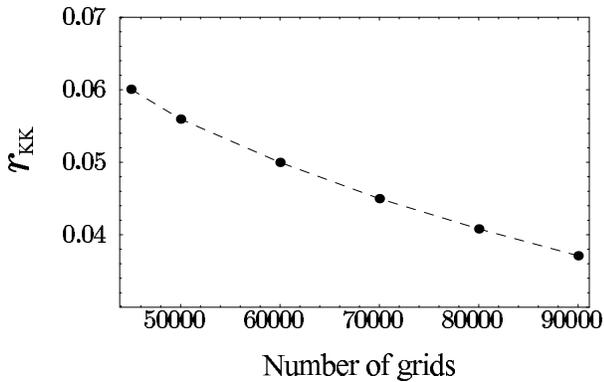}
  \end{center}
  \caption{$r_{{\rm KK}}$ for $\ell H_i=100$ and $k=13.3\times k_*$,
  versus the number of grids in the $U$-direction.
  }%
  \label{fig:converge.eps}
\end{figure}

Unfortunately, due to the limited number of
grids in the $U$-direction, it is difficult to
evaluate accurate values of $r_{{\rm KK}}$;
the convergence is not so good~(Fig.~\ref{fig:converge.eps}).
However, since $r_{{\rm KK}}$ decreases with an increasing number of grids,
it is strongly expected that
the effect of the initial KK fluctuations is negligibly small.
To obtain a more accurate evaluation of the contribution of
the initial KK modes,
we need an improved numerical formulation,
though it seems quite unlikely that
$r_{{\rm KK}}$ turns to increase at a much larger number of grids.
(We confirmed that the convergence of the zero mode part $\cP_0$ is
sufficiently good.)

\section{Generation of dark radiation}

So far we have concentrated on final zero mode gravitons
created from initial vacuum fluctuations.
In this section we shall discuss the generation of KK gravitons.
In particular, we are interested in the KK mode gravitons
created from initial fluctuations in the zero mode,
because
from the five-dimensional point of view
they are interpreted as gravitons that
escape from the brane.
At sufficiently late times,
all the emitted gravitons fall deep into the bulk,
and then the bulk spacetime is described as an
AdS-Schwarzschild black hole,
the mass of which divided by $a^4$
is viewed as the ``dark radiation'' from a brane
observer~\cite{Hebecker:2001nv, Langlois:2002ke, Langlois:2003zb, Leeper:2003dd}.

Before going to the estimation of the energy density
of final KK mode gravitons,
first let us take a look at the energy density
of zero mode gravitons $\rho_{{\rm GW}}$.
Since at low energies gravitational waves evolve
in a standard manner, their energy density
behaves as $\rho_{{\rm GW}}\propto a^{-4}$.
Therefore the ratio $\rho_{{\rm GW}}/\rho_r$
is an invariant quantity in the low energy regime,
\textit{irrespective of the cosmic expansion}.
Evaluating it at the end of the radiation stage,
we have
\begin{eqnarray}
\frac{\rho_{{\rm GW}}}{\rho_{r}}=\frac{\rho_{{\rm GW},0}}{\rho_{r,0}}
\simeq \ln\left(\frac{k_i}{k_0}\right)\cdot
\frac{\delta_T^2}{6}
\sim\delta_T^2.\label{rho_GW0}
\end{eqnarray}
Note that $\delta_T<10^{-5}$.
In deriving the estimate~(\ref{rho_GW0}),
we used the formula
\begin{eqnarray}
\rho_{{\rm GW}, 0} = M_{{\rm Pl}}^2\int_{k_0}^{k_i}
\!\!d\ln k~k^2\cP(k),
\label{en_gw}
\end{eqnarray}
and substituted the numerical result obtained in the previous section,
$\cP \simeq (\delta_T^2/2)(k/k_0)^{-2}$, where $k_0$
can be eliminated in favor of $\rho_{r,0}$
by using the Friedmann equation at low energies
$
k_0^2=H_0^2\simeq \rho_{r,0}/(3M_{{\rm Pl}}^2).
$
The upper limit of the integral may be given by
the inverse horizon scale at the end of inflation,
\begin{eqnarray}
k_i:=a_iH_i,
\end{eqnarray}
because the particle production
is exponentially suppressed on sub-horizon scales.

Let $\tilde\rho_{{\rm GW}}$ be the energy density
of gravitational waves
obtained by neglecting the mode mixing effect.
More precisely, $\tilde \rho_{{\rm GW}}$ is the energy density of
gravitational waves $h_{\mu\nu}$
where $h_{\mu\nu}$ is a solution of the conventional perturbation
equation $(\partial^2_{\tau}+3H\partial_{\tau}+k^2/a^2)h_{\mu\nu}=0$
with the cosmic expansion given by a solution of the
\textit{modified} Friedmann equation.
This would be much greater than $\rho_{{\rm GW}}$.
Then, $\Delta \rho:=\tilde\rho_{{\rm GW}}-\rho_{{\rm GW}}\simeq\tilde\rho_{{\rm GW}}$
is the energy density that leaks from the brane,
and by definition $\Delta\rho$ is proportional to $a^{-4}$
as long as it is evaluated in the low energy regime.
Thus, $\Delta\rho/\rho_r$
is an invariant quantity.
Now $\Delta\rho_{,0}$ can be calculated from
the spectrum of the form~(\ref{spectrum_incorrect}),
and we have an estimate
\begin{eqnarray}
\frac{\Delta\rho}{\rho_r}=\frac{\Delta \rho_{,0}}{\rho_{r,0}}\simeq
\frac{\delta_T^2}{8}\times \ell H_i,
\label{lost_energy}
\end{eqnarray}
where we used $a^4\rho_r=a_*^4\sigma=\rho_{r,0}$ and
the Friedmann equation at high energies,
$H^2\simeq\rho^2/6M_{{\rm Pl}}^2\sigma$.
The estimate~(\ref{lost_energy}) implies that
a large amount of energy (compared to $\rho_{{\rm GW}}$)
is lost from the brane.
Is the escaped energy $\Delta\rho$ directly transferred to
the final bulk gravitons?
To discuss this point,
we compare it with the energy density of the generated dark radiation.

Since KK modes are excited dominantly at high energies
but not at low energies,
the dark radiation, as is deduced from its name,
behaves like a radiation component, $\rho_{{\rm DR}}\propto a^{-4}$,
at late times.
Hence, we shall see the ratio $\rho_{{\rm DR}}/\rho_r$
but it may be evaluated at the end of the radiation stage.

The total number of the created bulk gravitons is given
again by the Wronskian as
\begin{eqnarray*}
&&\int dm\langle 0|\hat A_m^{\dagger}\hat A_m|0\rangle=
\\
&&\qquad\int dm\left[|(\phi_0^*\cdot \Phi_m)_i|^2
+\int d\nu|(\phi_{\nu}^*\cdot \Phi_m)_i|^2\right],
\end{eqnarray*}
where $\Phi_m$ is a solution of the Klein-Gordon equation~(\ref{KG_in_UV})
whose final configuration is a KK mode function $\varphi_m$
in the Minkowski phase.
Concentrating on the first part,
$|(\phi_0^*\cdot \Phi_m)|^2dm$
is identified as
the number of KK gravitons
coming from the initial zero mode fluctuations,
with their mass between $m$ and $m+dm$.
Thus the energy density is expressed as
\begin{eqnarray}
\rho_{{\rm DR},0}&:=&\int \frac{d^3k}{(2\pi)^3}\int dm~
\omega |(\phi_0^*\cdot \Phi_m)_i|^2
\nonumber\\
&=&\int d\ln k \int d\ln m~ \frac{mk^3\omega}{2\pi^2}
|(\phi_0^*\cdot \Phi_m)_i|^2.
\end{eqnarray}
As a side remark, in the junction model of Ref.~\cite{Gorbunov:2001ge},
the second part, i.e., the number of KK gravitons created
from the initial fluctuations in KK modes, is
suppressed compared to the first one.

\begin{figure}[t]
  \begin{center}
    \includegraphics[keepaspectratio=true,height=50mm]{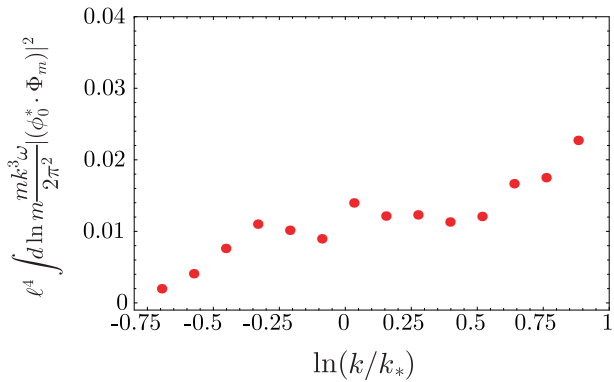}
  \end{center}
  \caption{Energy of bulk gravitons multiplied by the phase space factor.}%
  \label{fig:DR.eps}
\end{figure}

We numerically calculated the Wronskian $|(\phi_0^*\cdot \Phi_m)|^2$
in a similar manner to the previous section.
The Hubble parameter during inflation is chosen to be
$\ell H_i =42$.
In this case, the number of grids are
150000 in the $U$-direction and 20000 in the $V$-direction,
and the grid separation is about $0.0036\times \ell$.

The integrand
\begin{eqnarray*}
\int d\ln m~ \frac{mk^3\omega}{2\pi^2}
|(\phi_0^*\cdot \Phi_m)_i|^2
\end{eqnarray*}
is plotted in Fig.~\ref{fig:DR.eps}.
For each $k$, integration over $m$ is performed
up to $m\sim 2/\ell$ with a grid spacing of $\Delta \ln m \simeq 0.17$.
Performing the integration over $k$,
we obtain
\begin{eqnarray}
\rho_{{\rm DR},0}\approx 0.04\times\ell^{-4},
\label{dr-DR0}
\end{eqnarray}
where one should note that contributions from modes with $k<k_*$
are suppressed.
The radiation energy density can be written as
\begin{eqnarray}
\rho_{r,0}=a_i^4\rho_{r,i}
\simeq a_i^4(6M_{{\rm Pl}}^2\sigma)^{1/2}H_i
\simeq \frac{9}{2\pi^2}\frac{1}{\delta_T^2}(a_i H_i)^4,
\label{dr-r0}
\end{eqnarray}
which can be obtained by using the modified Friedmann equation and noting that
$\delta_T^2\simeq 3\ell H_i^3/4\pi^2M_{{\rm Pl}}^2$
for $\ell H_i\gg 1$.
From Eqs.~(\ref{dr-DR0}) and~(\ref{dr-r0}) we have an estimate
\begin{eqnarray}
\frac{\rho_{{\rm DR}}}{\rho_{r}}
< {\cal O}(1)\times \delta_T^2.
\label{generated_DR}
\end{eqnarray}
This result indicates that
the energy density of the generated dark radiation
is not larger than that of zero mode gravitons~[Eq.~(\ref{rho_GW0})].
Of course, this is a completely harmless amount of
an extra radiation component~\cite{Ichiki:2002eh}.
The scattering of particles on the brane in the early universe,
discussed in~\cite{Hebecker:2001nv, Langlois:2002ke, Langlois:2003zb, Leeper:2003dd},
can be a more efficient way
to produce bulk gravitons.


Although a large amount of energy is lost from
the brane via excitation of KK modes in the high energy regime,
the final energy density of the dark radiation
is much smaller than that, without an enhancement factor
like $\ell H_i$ in Eq.~(\ref{lost_energy}). 
This discrepancy is explained as follows~\cite{Hebecker:2001nv, Langlois:2003zb}.
In the high energy regime, the motion of the brane is so relativistic
(in the frame defined by the static bulk coordinates)
that emitted gravitons run almost parallel to the brane trajectory.
These gravitons stay in the vicinity of the brane and
bounce off it many times during the high energy stage,
until eventually they are reflected by the non-relativistic brane
to fall off into the bulk.
During this process, the gravitons lose a large portion of their momentum
transverse to the brane because they repeatedly
hit the retreating brane.
This qualitatively accounts for the smallness of the final energy density of
the dark radiation.
To justify the above interpretation quantitatively,
a more rigorous analysis will be needed
in the direction of Refs.~\cite{Langlois:2003zb, Minamitsuji:2005xs},
which includes calculating the pressure to the brane
due to the effective energy-momentum tensor of
the bulk gravitational waves.

Here we should comment on the result
of the junction model obtained by Gorbunov \textit{et al}.~\cite{Gorbunov:2001ge}.
In terms of the power spectrum, their result is summarized as
\begin{eqnarray}
\cP\approx\left\{
{\displaystyle
\frac{\delta_T^2}{2},\qquad(k\ll k_*),
}
\atop {\displaystyle
\frac{\delta_T^2}{2}\frac{4}{(k/k_*)^2},\quad(k_*\ll k \ll k_i),
}
\right.
\end{eqnarray}
where $k_*=a_*/\ell$, $k_i=a_iH_i$, and $a_*=a_i(=1)$
because a de Sitter inflationary stage is directly joined to
a Minkowski phase in the junction model
(see also Appendix A of Ref.~\cite{Kobayashi:2003cn}).
From this we can estimate the energy density
that leaks from the brane as
\begin{eqnarray}
\Delta\rho \approx M_{{\rm Pl}}^2H_i^2\delta_T^2
\approx \rho_{e.i.}\delta_T^2\times\ell H_i,
\end{eqnarray}
where $\rho_{e.i.}$ is the energy density at the ``end of inflation''.
On the other hand,
according to Appendix D of Ref.~\cite{Gorbunov:2001ge},
the energy density of created KK gravitons is given by
\begin{eqnarray}
\rho_{{\rm DR}} \approx H_i^4 \approx \rho_{e.i.}\delta_T^2.
\end{eqnarray}
Thus, we find that $\Delta \rho\sim \rho_{{\rm DR}}\times\ell H_i$,
which is consistent with our present result.
[Note that in the junction model the energy density of final zero mode gravitons
is estimated as $\rho_{{\rm GW}}\sim (M_{{\rm Pl}}^2/\ell^2)\delta_T^2$
and hence $\rho_{{\rm GW}} \ll \rho_{e.i.}\delta_T^2\sim \rho_{{\rm DR}}$.]

\section{Summary}

We have examined the power spectrum of
the gravitational wave background in the cosmological scenario of
the Randall-Sundrum braneworld.
There are three possible ingredients
which may lead the power spectrum to a non-standard one:
the unconventional background expansion rate
due to the $\rho^2$ term in the Friedmann equation,
the excitation of KK modes during the radiation-dominated stage
at high energies,
and the effect of initial vacuum fluctuations in KK modes.
Previous estimates are based on
a rather simple toy model~\cite{Gorbunov:2001ge}
or numerical studies about the classical evolution
of perturbations, neglecting the
initial KK fluctuations~\cite{Hiramatsu:2003iz, Hiramatsu:2004aa}.
In the present analysis,
initial conditions are set in a quantum-mechanical manner
and hence the effect of the initial KK fluctuations is included.
Along the same line in Ref.~\cite{Kobayashi:2005jx},
we make use of the Wronskian formulation
to obtain the final amplitude of
the zero mode gravitational waves numerically.
We have found that the effect of initial KK vacuum fluctuations
are subdominant: $r_{{\rm KK}}<0.1$.
Our result confirms that the damping of the amplitude due to
the KK mode excitation
and
the enhancement due to the modification of the background expansion rate
mainly work,
but almost cancel each other.
Consequently, the power spectrum is basically
the same as the standard one obtained in conventional four-dimensional cosmology.
We believe that the cancellation between the two effects
is a phenomenon peculiar to the radiation-dominated phase.
To make the particularity of the radiation stage clear,
it would be interesting to investigate
consequences of a different equation of state parameter $w(=p/\rho)$
after the inflationary stage.
This is the next issue we plan to report in a future publication.

Because of the limitation of our numerical computation,
we have not been able to give
a detailed evaluation of how much
the initial KK fluctuations contribute to the final power spectrum.
Although it is strongly indicated that
the initial KK effect never becomes larger than
the present evaluation,
to show that it is indeed true,
we need to improve the numerical formulation.
We will also report on this issue in a forthcoming paper.

We have also estimated the energy density of
the generated dark radiation numerically,
and shown that only a tiny amount is generated.
It is smaller than the energy density of zero mode gravitational waves.


\acknowledgments
This work was supported in part by Monbukagaku-sho Grants-in-Aid
for Scientific Research, Nos. 14047212, 16740165 and 17340075,
and by the Grant-in-Aid for the 21st Century COE
``Center for Diversity and Universality in Physics''
from the Ministry of Education, Culture, Sports, Science and Technology (MEXT)
of Japan.
T.K. is supported by the JSPS under Contract No.~01642.




\end{document}